\documentclass[preprint,showkeys,secnumarabic,amsfonts,showpacs,amsmath,amssymb]{revtex4}
\usepackage[dvips]{color}
\usepackage{array}
\usepackage{rotating}
\usepackage{epsfig}
\usepackage{amsmath}
\usepackage{graphicx}

\begin{document}
\title{ Stability Analysis in Tachyonic Potential Chameleon cosmology}

\author{H. Farajollahi}
\email{hosseinf@guilan.ac.ir} \affiliation{Department of Physics,
University of Guilan, Rasht, Iran}
\author{A. Salehi}
\email{a.salehi@guilan.ac.ir} \affiliation{Department of Physics,
University of Guilan, Rasht, Iran}
\author{F. Tayebi}
\email{ftayebi@guilan.ac.ir} \affiliation{Department of Physics,
University of Guilan, Rasht, Iran}
\author{A. Ravanpak}
\email{aravanpak@guilan.ac.ir} \affiliation{Department of Physics,
University of Guilan, Rasht, Iran}
\date{\today}

\begin{abstract}
 \noindent \hspace{0.35cm}
 We study general properties of attractors for tachyonic potential chameleon scalar-field model which possess
cosmological scaling solutions. An analytic formulation is given to
 obtain fixed points with a discussion on their stability. The model predicts a dynamical equation of state parameter with phantom crossing behavior for an accelerating universe. We constrain the parameters of the model by best fitting with the
recent data-sets from supernovae and simulated data points for redshift
drift experiment generated by Monte Carlo simulations.

\end{abstract}


\maketitle

\section{introduction}

Recently, the observations of high redshift type Ia supernovae and the surveys of clusters
of galaxies \cite{Tonry} reveal the universe accelerating expansion and that the density of matter
is very much less than the critical density. Also the observations of Cosmic Microwave
Background (CMB) anisotropies indicate that the universe is flat and the total energy density
is very close to the critical one \cite{Spergel}. While the above observational data properly complete each other,
 the dynamical dark energy proposal as an interesting possibility may arise to explain the observational constraints \cite{Sahni1}.

The two fine tuning problem and cosmic coincidence problem are the most serious issues with regards to the dark energy (DE) models and
of the most frequently used approach to moderate these problems is
the tracker field DE scenario by employing scalar field models which exhibit scaling solutions, for example see \cite{Easson}. Tracker models are independent of initial conditions used for field evolution but do
require the tuning of the slope of the scalar field potential. During the scaling regime, the
scalar field energy density is of the same order of magnitude as the background energy density. The scaling solutions
as dynamical attractors can considerably resolve the two above mentioned problems. Furthermore,
by investigating the nature of scaling solutions, one can determine whether such behavior
is stable or just a transient feature and explore the asymptotic behavior of the scalar field
potential \cite{Kim}.

In studying the canonical scalar field models with exponential potential there exist scaling attractor solutions \cite{Copeland}.
 In quintessence dark energy model, there are two scaling solutions. The first one
is fluid-scalar field scaling solution, which remains subdominant for most of the cosmic
evolution. It is necessary that the scalar field mimics the background energy density (radiation/
matter) in order to respect the nucleosynthesis constraint and can also solve the
fine-tuning problem of initial conditions. The second one is scalar field dominated scaling solution,
which is a late time attractor and gives rise to the accelerated expansion. Since the
fluid-scalar field scaling solution is non-accelerating, we need an additional mechanism exit
from the scaling regime so as to enter the scalar field dominated scaling solution at late
times. For the discussion on the exiting mechanism one can refer to \cite{Copeland1}.

In scalar-tensor theories \cite{Sahoo}--\cite{Nojiri3}, interaction of the scalar field with matter ( for example in chameleon cosmology) \cite{Setare1}--\cite{Dimopoulos} and the presence of the tachyon potential in the formalism \cite{Damouri}--\cite{Padmanabhan1} separately are used to interpret the late time acceleration. From a chameleonic point of
view, since the field mimics the background radiation/matter field, subdominant for most
of the evolution history except at late times when it becomes dominant, it may be regarded
as a cosmological tracker field \cite{faraj1}. For tachyon dark energy \cite{Tsujikawa,Piazza} as well as other scalar-tensor models such as chameleonic Brans-Dicke cosmology \cite{farajollahi}, the scaling solutions have been investigated separately.

Both Chameleon and
tachyon fields have been motivated separately from modified gravity and string theory \cite{chameleon}--\cite{tachyon} and explain the quintessence cosmological scenario and the late time universe acceleration. However in order to interpret also the phantom behavior of dark energy, we integrate both models into one to investigate in addition to the scaling solutions and late time acceleration of the universe, the phantom crossing and attractor behavior of the solutions which are also obtained in the context of non-minimally coupled scalar fields \cite{Carvalho} and are consistent with the observations.

The scalar field in our model plays the role of chameleon field coupled with the matter lagrangian and also by its presence in tachyonic potential can be regarded as tachyon field. The scalar field coupled with the matter lagrangian with a prefect fluid,  $p_{m}=\gamma \rho_{m}$, while the tachyonic scalar field candidates for dark energy.

The well-known geometric variables, i.e. Hubble and deceleration parameters
at the present time are used to explain the acceleration expansion of the universe. However,
considering the increased accuracy of the observational data during the last few years and
generality of the DE models, new geometrical variables are introduced to differentiate these
models and better fit the observational data. In this regard, the authors in \cite{Sahni} proposed a
cosmological diagnostic pair $\{s, r\}$, called statefinder, to differentiate the expansion dynamics
with higher derivatives of the scale factor and is a natural next step beyond the well known geometric variables. The
statefinder pair has been used to explore a series of dark energy and cosmological models,
including $\Lambda$ cold dark matter ($\Lambda$CDM), quintessence, coupled quintessence, Chaplygin gas, holographic dark energy
models, braneworld models, and so on \cite{Alam,Zimdahl,Yi}.

The "Cosmological
Redshift Drift" (CRD) test which maps the expansion of the universe
directly is also examined in here. We assume that the universe is
homogeneous and isotropic at the cosmological scales \cite{Lis}. The test is based on very simple and straightforward
physics. Observationally, it is a very challenging task and
requires technological breakthroughs, for more details see \cite{Cristiani}--\cite{Loeb}. we also verify our model by comparing the
 distance modulus measurements
versus redshift derived in the model with the data obtained from the observations of type
Ia supernovae and utilizing the $\chi^2$ method.

In the following, we study the detailed evolution of the model and the attractor
property of its solution. In Sec. 2, we derive the field equations. In Sec. 3, we obtain the autonomous equations of the model
and by using the phase plane analysis qualitatively investigate the dynamic of the
system and the existence of a late time attractor solution. We also examine the
behavior of the EoS parameter of the model and also perform a statefinder diagnostic for
the model and analyze the evolving trajectories of the model in the statefinder parameter
plane. Section 4 is devoted to perform cosmological tests. In sec. 5 we present summary and remarks.

\section{The Model}

The chameleon gravity with a tachyonic potential is given by,
\begin{eqnarray}\label{action}
S=\int[\frac{R}{16\pi
G}-V(\phi)\sqrt{1-\partial_\mu \phi \partial^\mu \phi }+f(\phi)\mathcal{L}_{m}]\sqrt{-g}dx^{4},
\end{eqnarray}
where $R$ is Ricci scalar, $G$ is the newtonian constant gravity, and the second term in the action is tachyon potential.
 Unlike the usual Einstein-Hilbert action in chameleon cosmology, the matter
Lagrangian $\mathcal{L}_{m}$ is modified as $f(\phi)\mathcal{L}_{m}$ , where $f(\phi)$ is
an analytic function of the scalar field. The last term in Lagrangian brings about the nonminimal
interaction between the matter and chameleon field.
The variation of action (\ref{action})  with respect to the metric tensor components in a spatially flat FRW  cosmology
yields the field equations:
\begin{eqnarray}
&&3H^{2}=\rho_{m}f+\frac{V(\phi)}{\sqrt{1-\dot{\phi}^{2}}},\label{fried1}\\
&&2\dot{H}+3H^2=-\gamma\rho_{m}f+V(\phi)\sqrt{1-\dot{\phi}^{2}},\label{fried2}
\end{eqnarray}
where we put  $8\pi G=c=\hbar=1$ and $ H=\frac{\dot{a}}{a}$  with $a$ is the scale factor of the universe. We also assume a perfect fluid with $p_{m}=\gamma\rho_{m}$. Note that in here $\gamma$ is the EoS parameter for nonminimally coupled chameleon field with the matter field in the universe. Variation of the action (\ref{action}) with respect to the scalar field  $\phi$ provides the wave
equation for chameleon field as
\begin{eqnarray}\label{phiequation}
\ddot{\phi}+(1-\dot{\phi}^{2})(3H\dot{\phi}+\frac{V^{'}}{V})=\frac{\epsilon f^{'}}{V}(1-\dot{\phi}^{2})^{\frac{3}{2}}\rho_{m},
\end{eqnarray}
where prime indicates differentiation with respect to $\phi$ and $\epsilon=\frac{1-3\gamma}{4}$.
From equations (\ref{fried1})--(\ref{fried2}) one arrives at the conservation equation,
\begin{eqnarray}\label{conserv2}
\dot{(\rho_{m}f)}+3H(1+\gamma)\rho_{m}f=-\epsilon \rho_{m}\dot{f}.
\end{eqnarray}

From equations (\ref{fried1}) and (\ref{fried2}), by defining the effective energy density and pressure,  $\rho_{eff}$ and $p_{eff}$, one can identify an effective EoS parameter as
\begin{eqnarray}\label{omegaef}
\omega_{eff}\equiv\frac{p_{eff}}{\rho_{eff}}=\frac{\gamma\rho_{m}f-V(\phi)\sqrt{1-\dot{\phi}^{2}}}{\rho_{m}f+\frac{V(\phi)}{\sqrt{1-\dot{\phi}^{2}}}}
\end{eqnarray}
In the next section we study the stability of the system in the frame work of phase-space trajectory analysis.

\section{STABILITY ANALYSIS-PERTURBATION AND PHASE SPACE}

The structure of the dynamical system can be studied via  phase plane analysis,
by introducing the following dimensionless variables,
\begin{eqnarray}\label{defin}
 x={\frac{\rho_{m}f}{3 H^{2}}},\ \ y=\frac{V}{3H^{2}},\ \ z=\dot{\phi},\ \ w=\frac{1}{H}.
\end{eqnarray}
We consider that $f(\phi)=f_{0}\exp{(\delta_{1}\phi)}$ and $V(\phi)=V_{0}\exp{(\delta_{2}\phi)}$ where $\delta_{1}$ and $ \delta_{2} $ are
dimensionless constants characterizing the slope of potential $V(\phi)$ and coupling field $f(\phi)$. The cosmological models with such exponential functions have been known lead to interesting physics
in a variety of context, ranging from existence of accelerated expansions \cite{Halliwell} to cosmological
scaling solutions \cite{Ratra}--\cite{Yokoyama}. In particular, the exponential forms of $f(\phi)$ and $V(\phi)$ are motivated by chameleon models \cite{chameleon} and also from stability considerations \cite{stability}. In addition, attractor solutions with exponential functions may lead to cosmic acceleration for natural values of model parameters \cite{Barreiro}.

By using (\ref{fried1})-(\ref{conserv2}), the equations for the new dynamical variables are,
\begin{eqnarray}
\frac{dx}{dN}&=&x[\frac{3yz^{2}}{\sqrt{1-z^{2}}}-\delta_{1}\epsilon wz], \label{x1}\\
\frac{dy}{dN}&=&y[3(1+\gamma)x+3\frac{yz^{2}}{\sqrt{1-z^{2}}}+\frac{\delta_{2}z}{w}], \label{y1}\\
\frac{dz}{dN}&=&(z^{2}-1)(3z+\delta_{2}w)+\epsilon\delta_{1}w(z^{2}-1)^{\frac{3}{2}}\frac{x}{y}, \label{z1}\\
\frac{dw}{dN}&=&w[\frac{3}{2}(1+\gamma)x+\frac{3}{2}\frac{yz^{2}}{\sqrt{1-z^{2}}}],\label{w1}
\end{eqnarray}
where $N = ln$(a). By using the Fridmann constraint equation (\ref{fried1}) in terms of the new dynamical variables:
\begin{eqnarray}\label{constraint}
x+\frac{y}{\sqrt{1-z^{2}}}=1,
\end{eqnarray}
the equations (\ref{x1})-(\ref{w1}) reduce to,
\begin{eqnarray}
\frac{dx}{dN}&=&x[3(1-x)z^{2}-\delta_{1}\epsilon wz], \label{x2}\\
\frac{dz}{dN}&=&(z^{2}-1)(3z+\delta_{2}w)+\frac{\epsilon\delta_{1}wx(z^{2}-1)}{1-x}, \label{z2}\\
\frac{dw}{dN}&=&w[\frac{3}{2}(1+\gamma)x+\frac{3}{2}(1-x)z^{2}]. \label{w2}
\end{eqnarray}
In stability formalism, by simultaneously solving $x'=0$, $y'=0$ and $w'=0$, where prime from now on means derivative with respect to $ln (a)$,
 the fixed points (critical points) can be obtained.
The critical points that may explicitly depend on the cosmological and stability parameters $\gamma$, $\delta_{1}$ and $\delta_{2}$ are
illustrated in Table I. Substituting
linear perturbations $x'\rightarrow x'+\delta x'$, $z'\rightarrow z'+\delta z'$, $w'\rightarrow w'+\delta w'$ about the critical points into
the three independent equations (\ref{x2})--(\ref{w2}), to the first
orders in the perturbations, gives us three eigenvalues $\lambda_{i} (i=1,2,3)$ which has to be negative as a requirement by stability method.
\begin{table}[ht]
\caption{critical points} 
\centering 
\begin{tabular}{c c c c c c} 
\hline\hline 
points  &  P1  & P2 \ & P3 \ & P4 \ & P5  \\ [4ex] 
\hline 
$x$ & 0 & 0 & $x(t)$ & $-\frac{1}{\gamma}$ & $-\frac{1}{\gamma}$  \\ 
\hline 
$z$ & 1 & -1 & 0 & 1 & -1  \\
\hline 
$w$ & 0 & 0 & 0 & $\frac{-12(1+\gamma)}{\delta_{1}\gamma(-1+3\gamma)}$ & $\frac{12(1+\gamma)}{\delta_{1}\gamma(-1+3\gamma)}$ \\
\hline 
\end{tabular}
\label{table:1} 
\end{table}

In the following, the nature of the five critical points are given with the stability conditions as

\begin{eqnarray}\label{p15}
P1&:& \lambda_{1P1}=6,\lambda_{2P1}=3,\lambda_{3P1}=\frac{3}{2},unstable\nonumber\\
P2&:& \lambda_{1P2}=6,\lambda_{2P2}=3,\lambda_{3P2}=\frac{3}{2},unstable\nonumber\\
P3&:& \lambda_{1P3}=0,\lambda_{2P3}=-3,\lambda_{3P3}=\frac{3}{2}x(t)(1+\gamma),unstable\nonumber\\
P4&,& P5: \lambda_{1P4,5}= \frac{3(1+\sqrt{1+2\gamma^{2}+2\gamma})}{2\gamma}, \lambda_{2P4,5}=\frac{-3(-1+\sqrt{1+2\gamma^{2}+2\gamma})}{2\gamma},
\nonumber\\
&&\lambda_{3P4,5}=\frac{6 (2 \delta_{1} \gamma+3 \delta_{1} \gamma^{2}-4 \delta_{2}-4 \delta_{2} \gamma-\delta_{1}) }{\gamma (-1+3\gamma)\delta_{1}}
,\nonumber\\
&& \mbox{stable for} \delta_{2} < \frac{1}{4}(3\gamma-1)\delta_{1} , \ \  -1 <\gamma<0, \ \  \frac{-24(1+\gamma)}{\gamma \delta_{1}(3\gamma-1)}>0,\\ \nonumber
&& \ \ \mbox{\ \ \ \ or} \ \ \ \  \delta_{2} > \frac{1}{4}(3\gamma-1)\delta_{1} ,\ \   -1 <\gamma<0, \ \
 \frac{-24(1+\gamma)}{\gamma \delta_{1}(3\gamma-1)}<0
\end{eqnarray}

As can be seen, the critical points P1, P2 and P3 are unstable and the points P4 and P5 are stable for the given conditions on  $\delta_{1}$, $\delta_{2}$
and $\gamma$. It is interesting to note that in order for the critical points p4 and P5 to be stable, in addition to the conditions on the stability parameters $\delta_{1}$ and $\delta_{2}$, the EoS parameter of the matter field has to be between zero and $-1$. This means that the system is stable for the chameleon field interaction with dark energy. In the following we will numerically investigate the models
with specific choice of EoS parameter of the matter in the universe. In Fig. 1, the attraction of the trajectories to the
critical points, P4 and P5 in the 3-dim phase plane is shown for the given initial conditions.

\begin{tabular*}{2.5 cm}{cc}
\includegraphics[scale=.4]{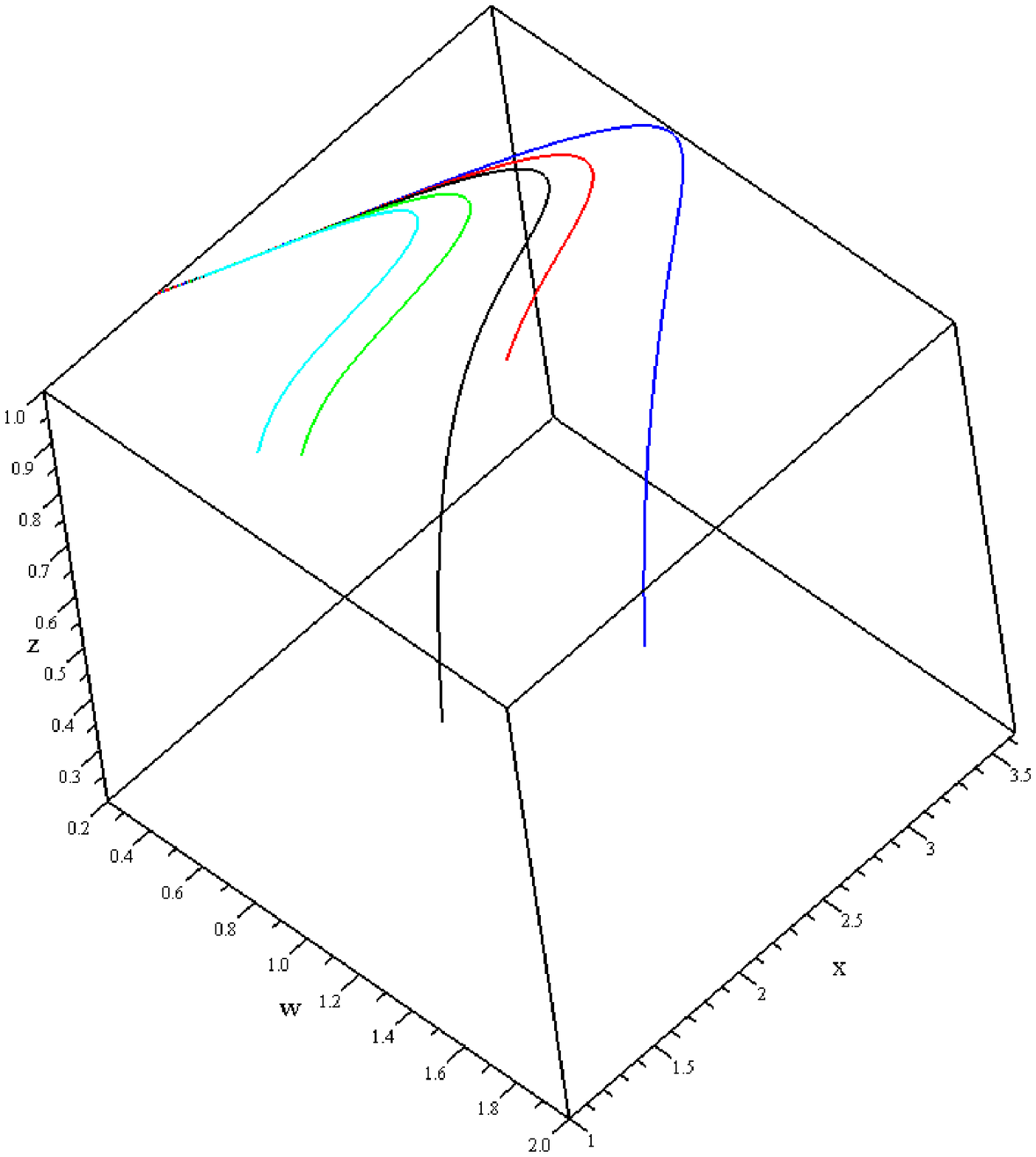}\hspace{0.1 cm}\includegraphics[scale=.45]{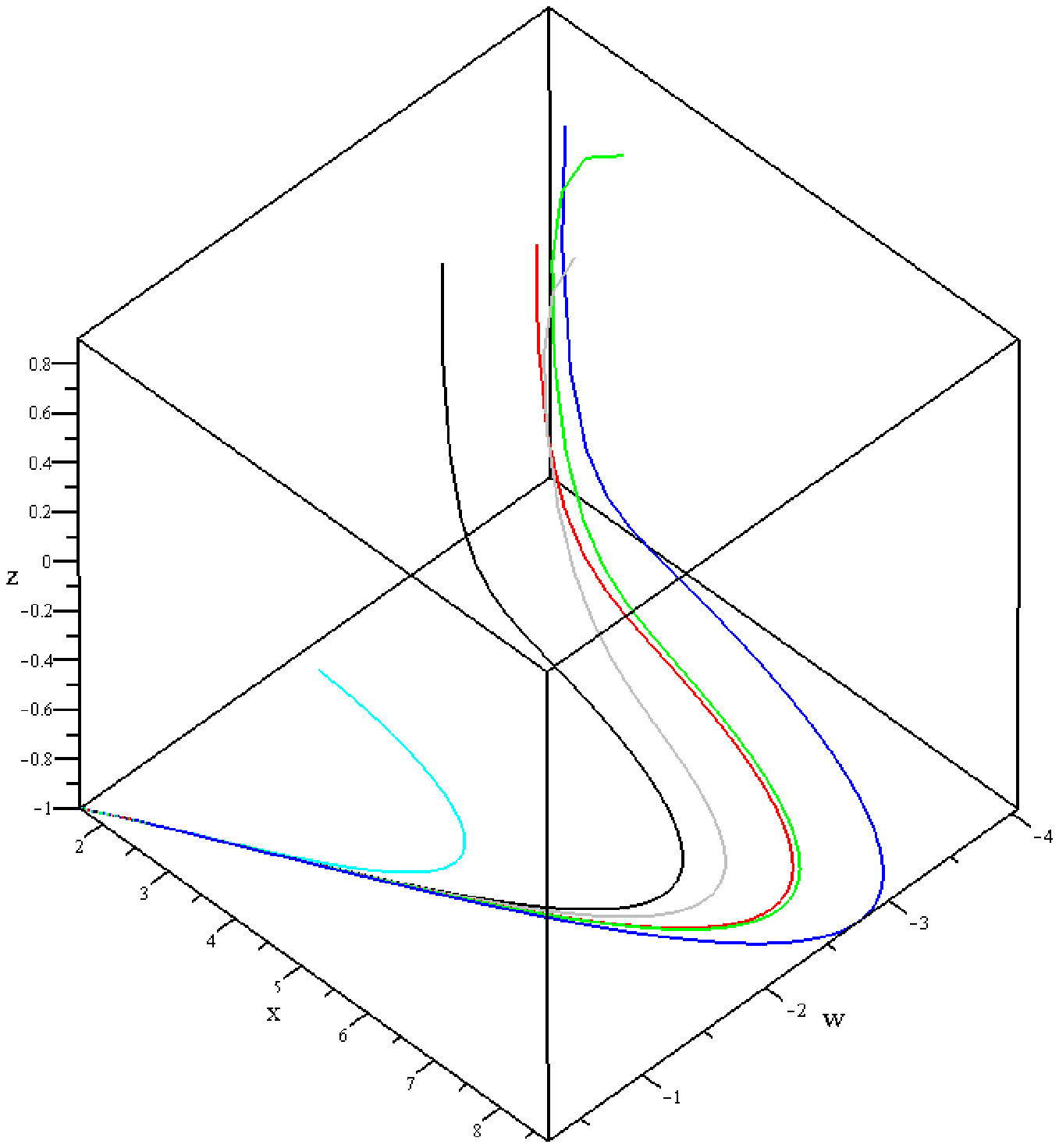}\hspace{0.1 cm}\\
Fig. 1:  The attractor property of the dynamical system in the 3-dim phase plane\\
 for the case $\gamma=-0.6$, $\delta_{1}=-12$ and $\delta_{2}=-10$ and different initial conditions
\end{tabular*}\\

Fig. 2 shows the projection of the 3-dim phase space of P4 and P5 onto the 2-dim phase plane:\\

\begin{tabular*}{2.5 cm}{cc}
\includegraphics[scale=.35]{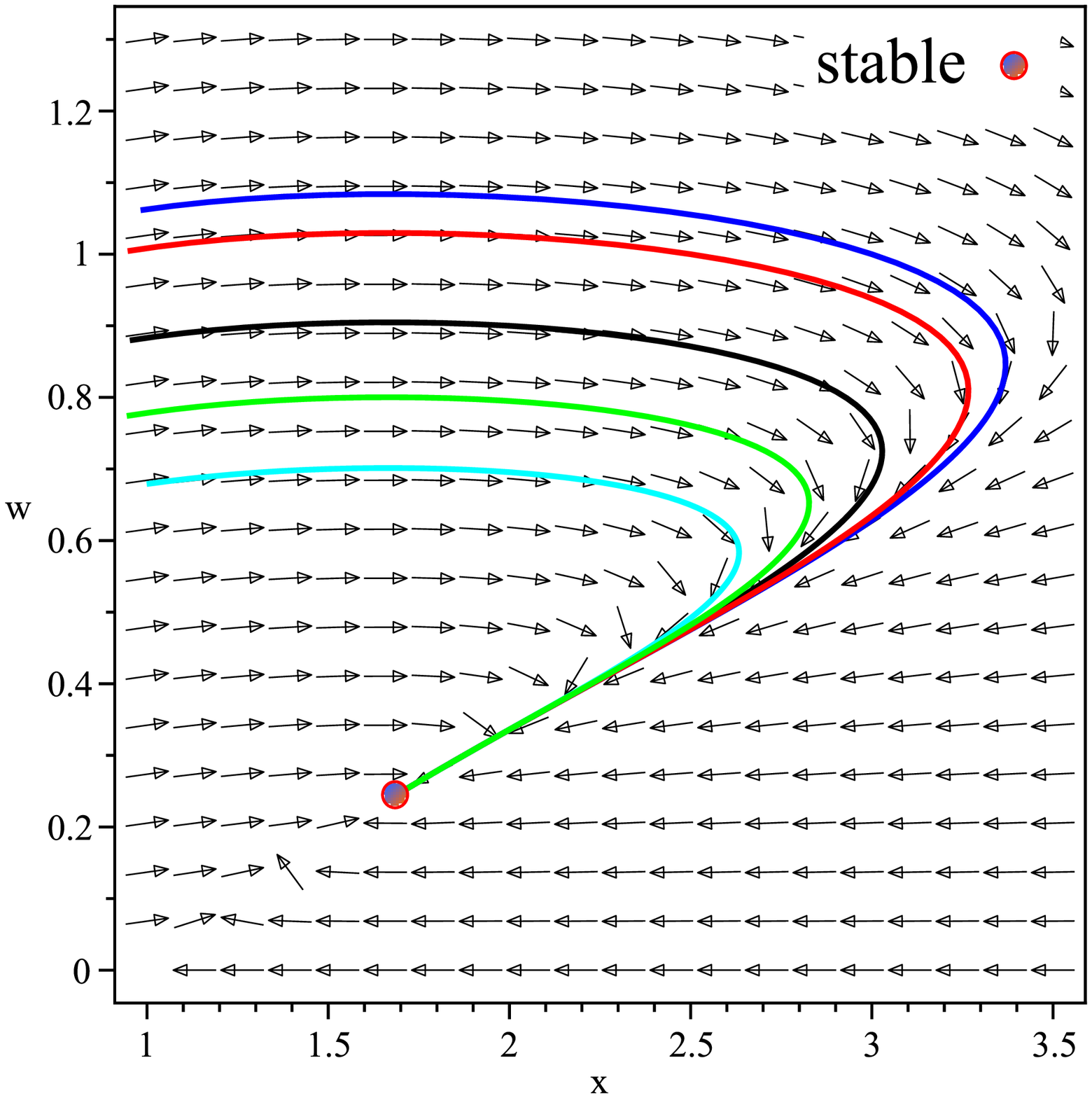}\hspace{0.1 cm}\includegraphics[scale=.35]{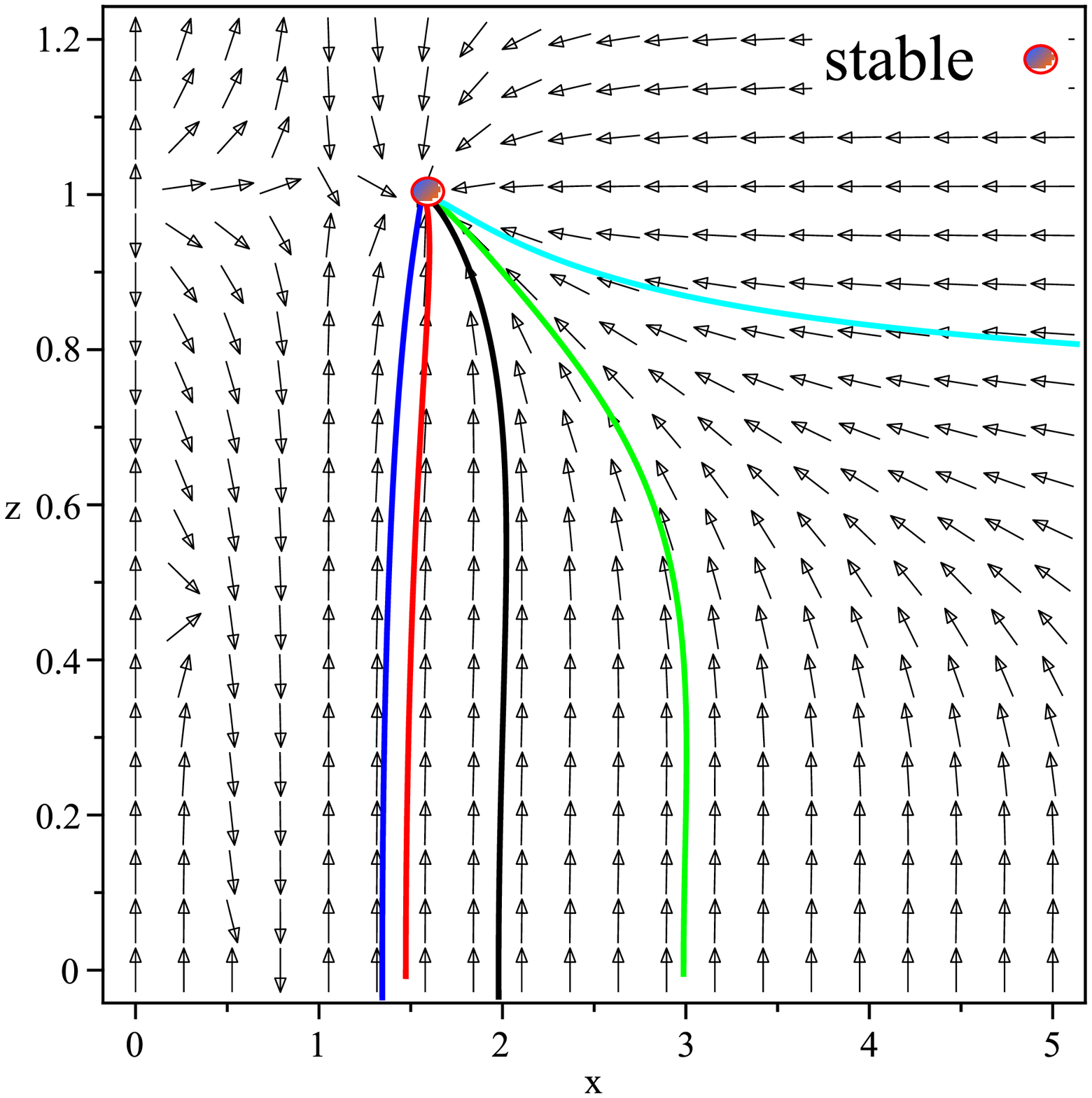}\hspace{0.1 cm}\\
\includegraphics[scale=.35]{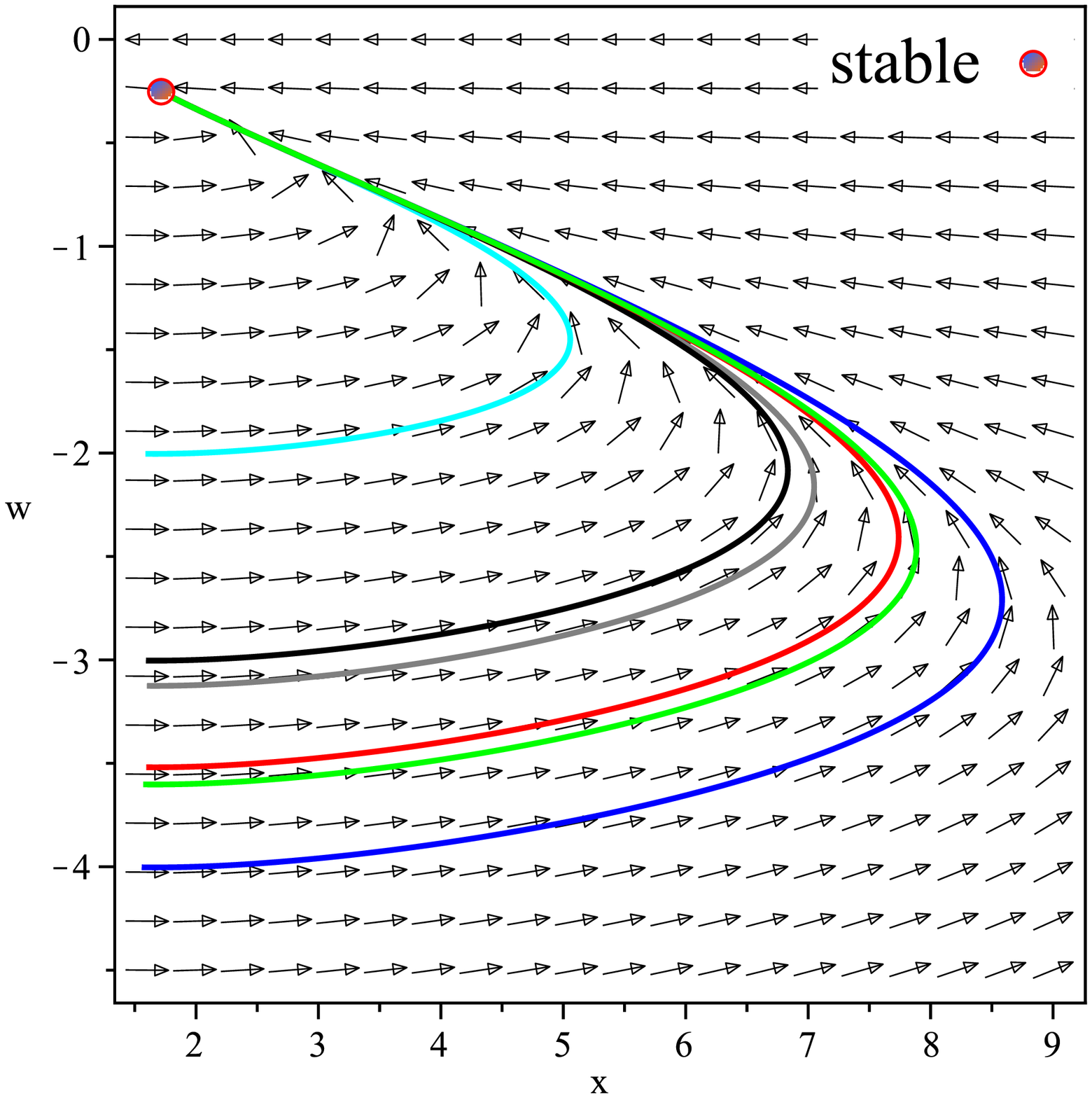}\hspace{0.1 cm}\includegraphics[scale=.38]{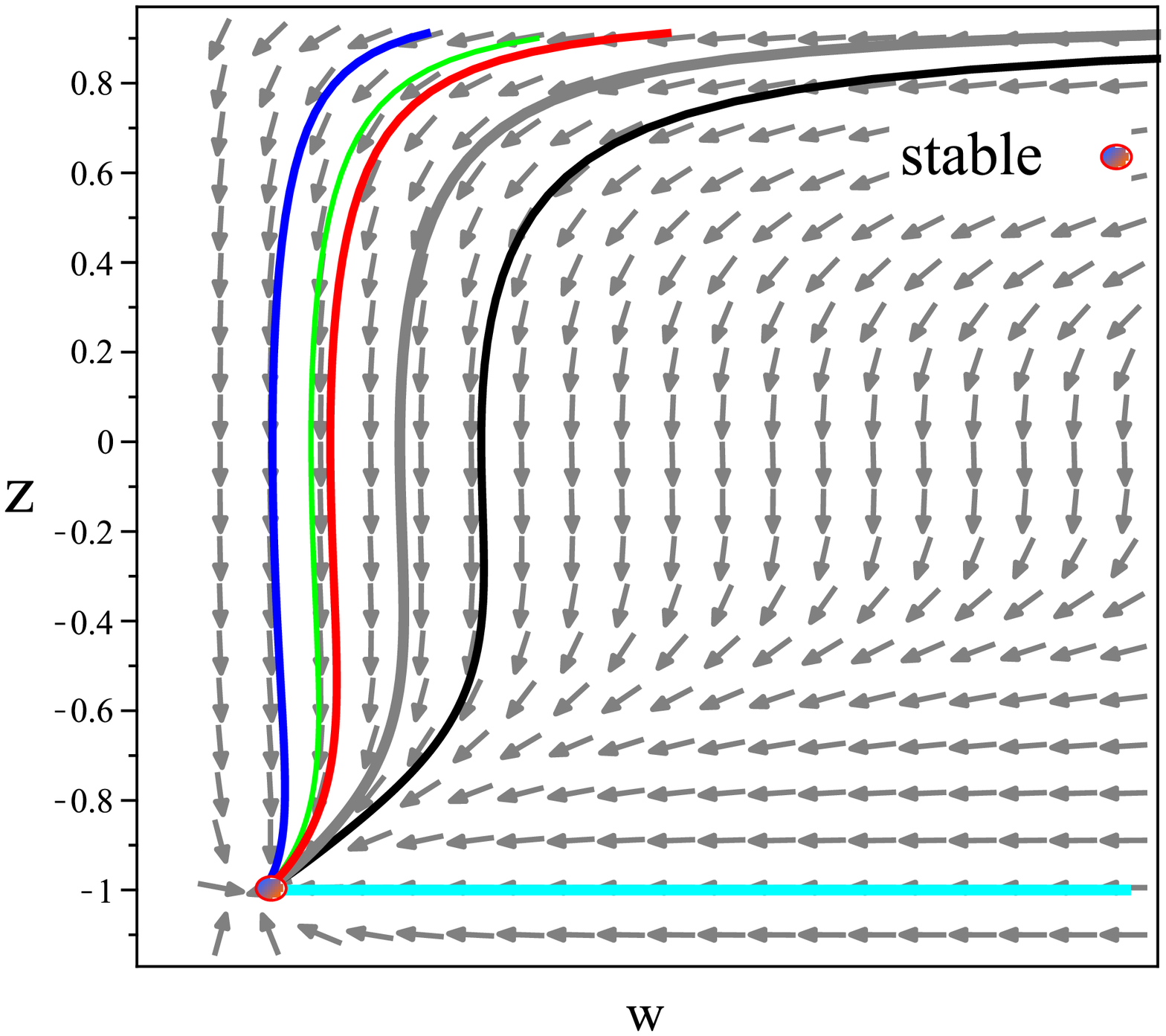}\hspace{0.1 cm}\\
Fig. 2: The 2-dim phase plane corresponding to the critical points P4(top figures)
and \\P5 (bottom figures). We have $\gamma=-0.6$, $\delta_{1}=-12$ and $\delta_{2}=-10$\\
{\bf I.C.s:top)left)}blue:$x(0)=2$,$w(0)=-4$,red:$x(0)=1.5$,$w(0)=2$,black:$x(0)=2$,$w(0)=0.9$\\
,green:$x(0)=2.8$,$w(0)=0.7$,cyan:$x(0)=2.5$, $w(0)=0.06$,{\bf top)right)}blue:  $x(0)=2$,\\
$z(0)=0.1$,red:$x(0)=1.5$,$z(0)=0.2$,black:$x(0)=2$,$z(0)=0.6$,green:$x(0)=2.8$,\\
$z(0)=0.6$,cyan:$x(0)=2.5$,$z(0)=0.9$,{\bf bottom)left)}blue:$x(0)=2$,$w(0)=-4$,red:$x(0)=2.5$,\\
$w(0)=-3.5$,black:$x(0)=2$,$w(0)=-3$,green:$x(0)=3.8$, $w(0)=-3.5$,cyan:$x(0)=2$,\\
$w(0)=-2$,gray:$x(0)=2.5$, $w(0)=-3$,{\bf bottom)right)}blue:  $x(0)=2$, $z(0)=0.3$,\\
red:$x(0)=1.5$, $z(0)=0.8$,black:$x(0)=2$,$z(0)=-0.9$,green: $x(0)=2.8$, $z(0)=0.9$,\\
cyan:  $x(0)=2.5$, $z(0)=-1$,gray:  $x(0)=2.5$, $x(0)=-0.9$,
\end{tabular*}\\

The particular choices of the model parameters $\delta_1$, $\delta_2$, $\gamma$ and initial conditions are just examples in expression (\ref{p15}) within the range of stable critical points P4 and P5 (the attractors). If we chose any other values for these parameters within the range, and initial conditions for the dynamical variables near the fixed points P4 and P5, we still get the same
behavior for the trajectories. That is, with small perturbation, the trajectories approach the same stable critical points as shown in figures 1 and 2.

Two of the cosmological parameters which relates the dynamics of the universe with the observational data are the EoS and deceleration parameters of the cosmological model. In terms of the new dynamical variables in our model they are given by,
\begin{eqnarray}
\omega_{eff}&=&\gamma x-y\sqrt{1-z^{2}}\label{w},\\
q&=&-1-\frac{\dot{H}}{H^2}=-1+\frac{3}{2}(\gamma x-y\sqrt{1-z^{2}}+1)\label{q}.
\end{eqnarray}
Moreover, the statefinder parameters $\{r,s\}$, the effective squared sound speed and the power $p$ in the scale factor $a\varpropto t^p$ in terms of the dynamical variables are,
\begin{eqnarray}
r&=&\frac{3}{2}z(2-z^{2})\mathcal{A} +\frac{3w}{2}(1-x)\mathcal{B} +\frac{9}{2}x\gamma(\gamma+1)-\frac{9}{2}(1-x)z^{2}+1,\label{r}\\
s&=&\frac{\frac{3}{2}z(2-z^{2})\mathcal{A} +\frac{3w}{2}(1-x)\mathcal{B}+\frac{9}{2}x\gamma(\gamma+1)-\frac{9}{2}(1-x)z^{2}}{\frac{9}{2}(\gamma x-y\sqrt{1-z^{2}})},\label{s}\\
c_s^2&=&\omega_{eff}-\frac{\omega'_{eff}}{3(\omega_{eff}+1)}=\frac{\gamma x'+x'(1-z^2)+2(1-x)zz'}{3((1-x)(1-z^2)-1-\gamma x)}+\gamma x'-(1-x)(1-z^2),\\
p&=&\frac{2}{3(\gamma x-y\sqrt{1-z^{2}}+1)},\label{P}
\end{eqnarray}
where $\mathcal{A} \equiv3(1-x)z+\delta_{2}w(1-x)-\delta_{1}\epsilon xw$ and $\mathcal{B}\equiv(1+\gamma)\epsilon\delta_{1}x\sqrt{1-z^{2}}-\delta_{2}z^{2}.$

For our model these parameters are presented in Table II,\\
\begin{table}[hb]
\caption{Properties of the critical points } 
\centering 
\begin{tabular}{c c c c c c c } 
\hline\hline 
points &  $c_s^{2}$  & $q$ & $\omega_{eff}$ & r & s & acceleration \\ [3ex] 
\hline 
P1   &0& 1/2 & 0 & $1$&$\infty$& No\\ 
P2  &0& 1/2 & 0 & $1$&$\infty$& No \\
P3 &$x(\gamma+1)-1$& $\frac{3}{2}x(\gamma+1)-1$ & $x(\gamma+1)-1$ &$1+\frac{9}{2}x\gamma(\gamma+1)$& $\frac{x\gamma(\gamma+1)}{x(\gamma+1)-1}$& Conditional\\
P4 &-$\infty$& -1 & -1 &$-\frac{7}{2}+\frac{9}{2}(\frac{1+\gamma}{\gamma^{2}}-\gamma)$ & $-\frac{1+\gamma}{\gamma^{2}}+\gamma+1$& Yes \\
P5 &-$\infty$& -1 & -1 &$-\frac{7}{2}+\frac{9}{2}(\frac{1+\gamma}{\gamma^{2}}-\gamma)-36\mathcal{C} $&
$8\mathcal{C}-\frac{1+\gamma}{\gamma^{2}}+\gamma+1 $ & Yes \\
 [1ex] 
\hline 
\end{tabular}
\label{table:2} 
\end{table}\\

where $\mathcal{C}$ is $\frac{8\delta_{2}(1+\gamma)(1+\frac{1}{\gamma})}{\delta_{1}\gamma(3\gamma-1)}$.

Note that with the disturbances propagate with the sound speed, the larger the sound
speed, larger the range of influence. As
long as the sound speed is finite, still the influence moves locally . For infinite sound speed, then disturbances
can propagate instantaneously, hence
the system becomes non-local and this is the case for incompressible fluids
($\delta\rho_{eff} = 0$) \cite{verma}. As can be seen from Table II and Fig.3, the universe starts from an unstable state (critical point P1 and P2) in the past when $\omega_{eff}=0$. It crosses the phantom divide line in near future at about $0.002<z<0.004$ and eventually become tangent to the phantom divide line at far future when approaches the stable critical point P5. Note that for the stable critical points P4 and P5, the squared sound speed is minus infinity as expected from the behavior of  $\omega_{eff}$ (Fig. 3) approaching phantom divide line in far future.

In our model, based on stability analysis, the critical points P4 and P5 are stable. That is, for a small perturbation of the parameters, the universe starts from the unstable state say P1 and approaches the stable state P4 or P5. However, since the stable critical points are phantom state with $\omega_{eff}=-1$, one would expect $c_s^2$ become minus infinity which is related to the nature of phantom energy.

Also, from Fig. 3, the universe  begins from an unstable state in the past with $\omega_{eff}=0$ and reaches a stable state in the future (phantom state) with $\omega_{eff}=-1$. Our result is in agreement with the work in \cite{Yang} where from stability point of view some of the fixed points in the cosmological model are stable whereas they possess negative value for $c_s^2$ \cite{Pedro}

\begin{tabular*}{2.5 cm}{cc}
\includegraphics[scale=.4]{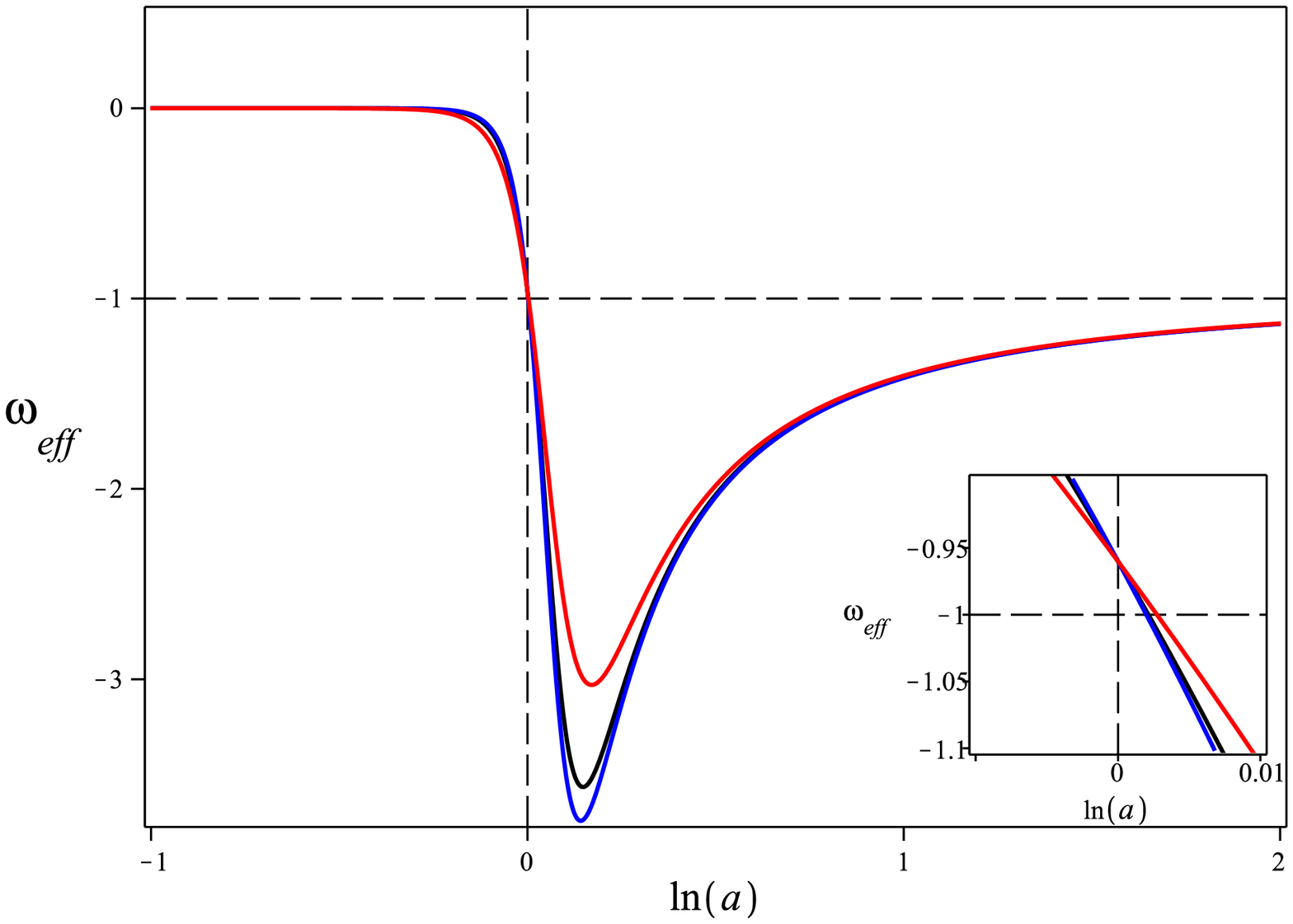}\hspace{0.1 cm}\\
Fig. 3: The evolution of EoS parameter vs $ln(a)$. ICs. $x(0)=1.6$, $z(0)=1$, $w(0)=2$, \\
$\epsilon=0.7$ and $\delta_{2}=-10$, (black): $\delta_{1}=-15$, (red): $\delta_{1}=-12$, (blue): $\delta_{1}=-16$.\\
\end{tabular*}\\

Fig.4 shows the statefinder diagrams $\{s,q\}$ and $\{r,q\}$ evolutionary
trajectories. From the graph of the statefinder $\{r,q\}$ all the trajectories
for different values of stability parameter $\delta_1$ start from the standard cold dark energy (SCDM) in the past which is an
unstable critical point and end their evolution to the critical point correspond to the state next to steady state (SS) in the
future. The current values of the trajectories are shown and can be compared with the
position of SCDM, LCDM and SS. From the graph of the statefinder $\{s,q\}$ one observes that all the
trajectories begin from the same unstable critical point and approach a state close to SS in the future.
As can be seen, while the current values of the s and q are shown in the graph, all trajectories approach the stable critical point which is an attractor in the future.

\begin{tabular*}{2.5 cm}{cc}
\includegraphics[scale=.4]{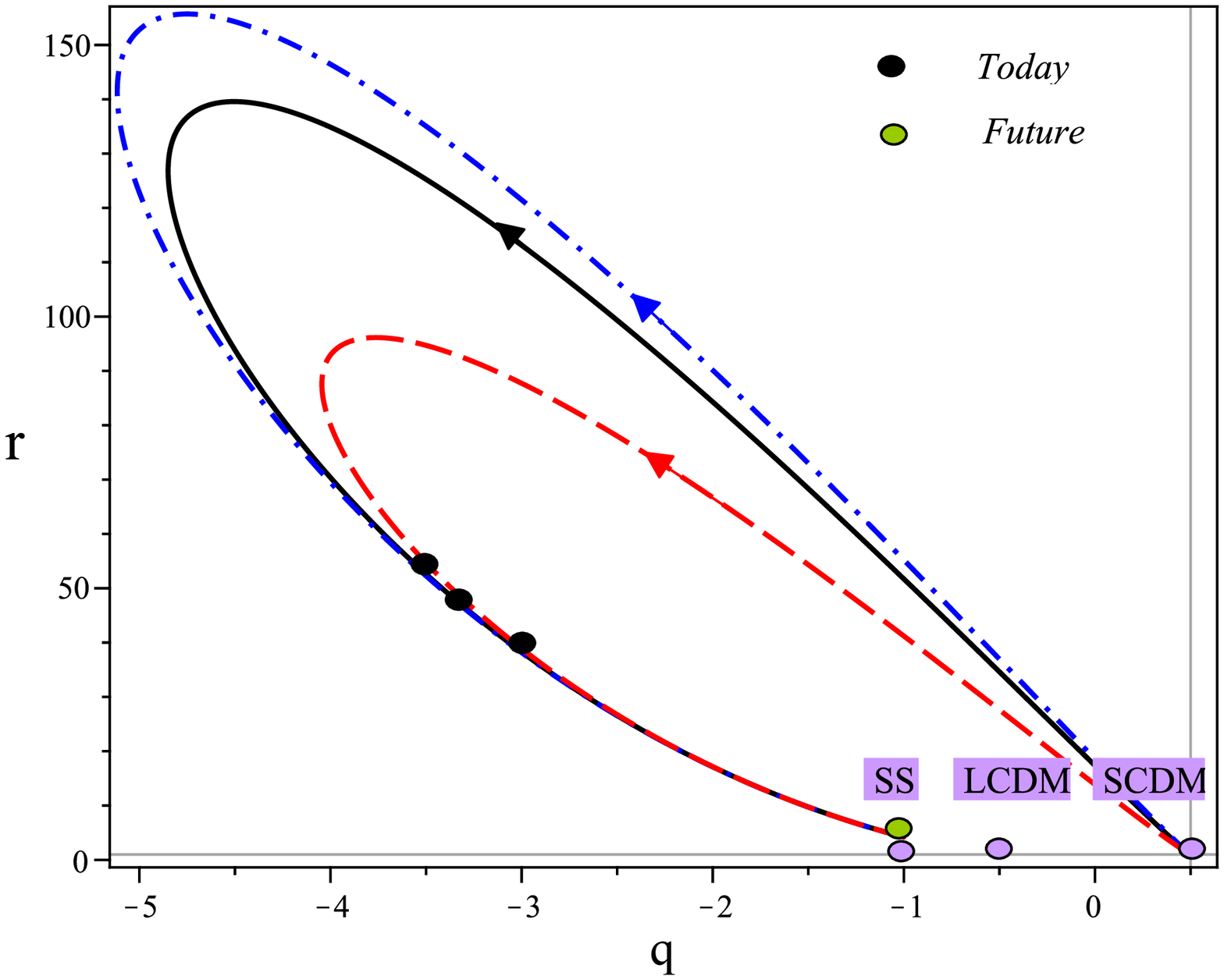}\hspace{0.1 cm}\includegraphics[scale=.4]{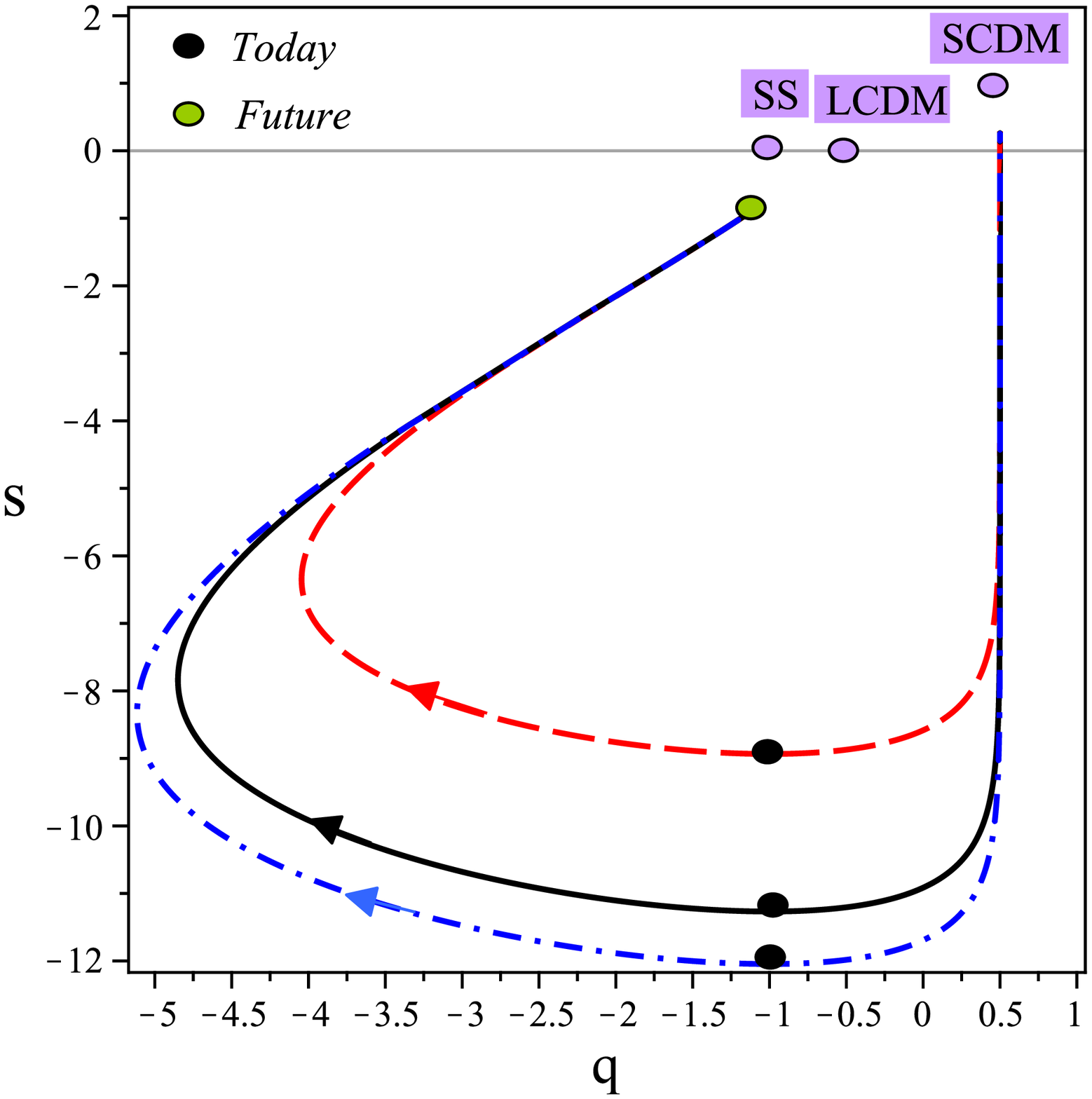}\hspace{0.1 cm}\\
Fig.4:  Trajectories in the statefinder plane $\{r, q\}$ and $\{s,q\}$. I.Cs. $x(0)=1.6$, $z(0)=1$,\\
 $w(0)=2$, $\epsilon=0.7$ and $\delta_{2}=-10$,
(black): $\delta_{1}=-15$, (red): $\delta_{1}=-12$, (blue): $\delta_{1}=-16$.\\
\end{tabular*}\\

In addition, Fig.5 shows the statefinder diagram $\{r,s\}$ evolutionary trajectory. From the graph we
see that all the trajectories for different values of $\delta_1$ commence evolving from different points
in the past and continue towards the same stable critical
point in the future near SS.

\begin{tabular*}{2.5 cm}{cc}
\includegraphics[scale=.4]{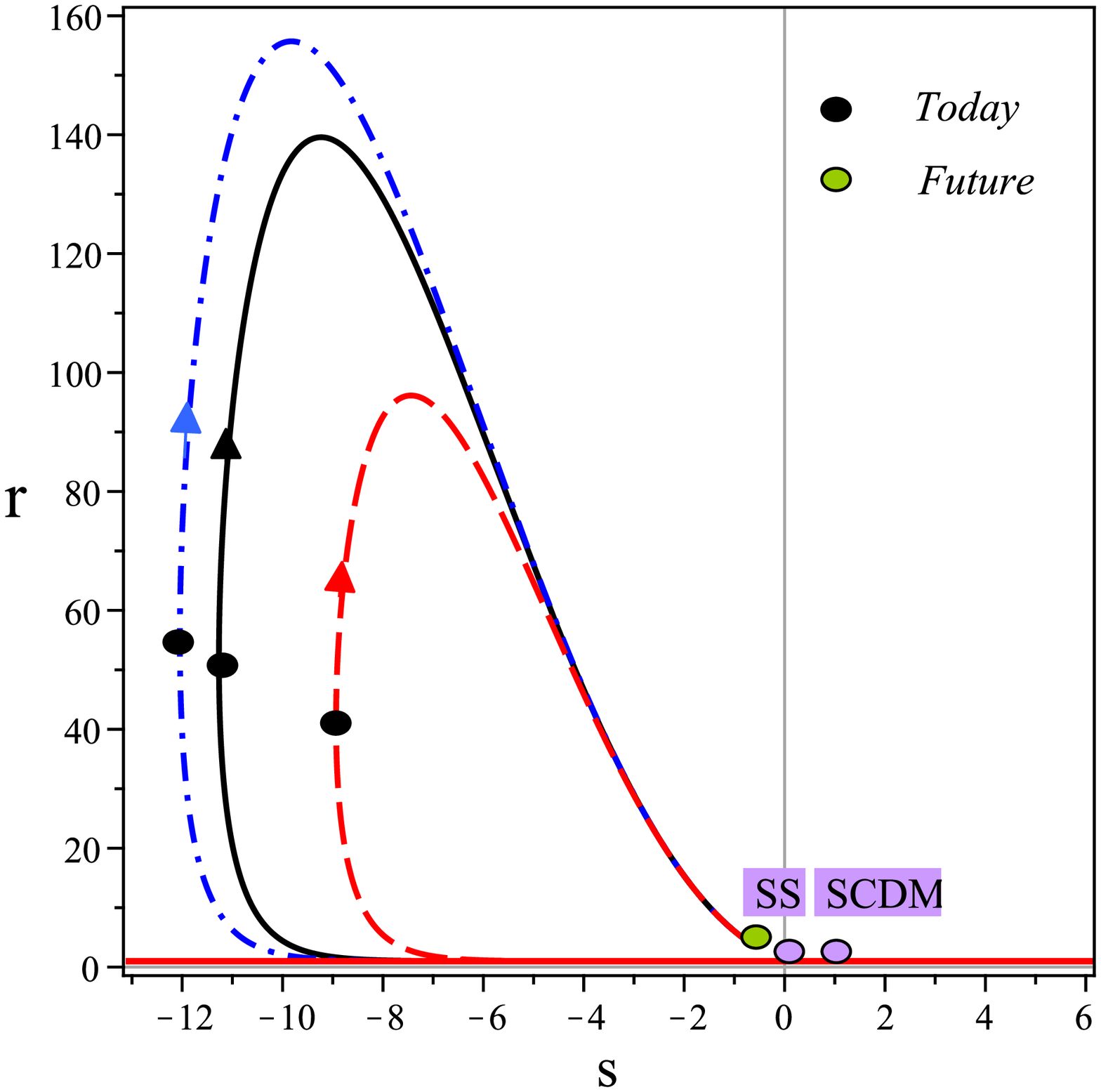}\hspace{0.1 cm}\\
Fig. 5: Trajectories in the statefinder plane \{r, s\}, I.Cs. $x(0)=1.6$, $z(0)=1$, $w(0)=2$,\\
 $\epsilon=0.7$ and $\delta_{2}=-10$, (black): $\delta_{1}=-15$, (red): $\delta_{1}=-12$, (blue): $\delta_{1}=-16$.\\
\end{tabular*}\\

In Figs. 3,4 and 5, for effective EoS parameter and statefinders, we have selected stability parameters that are within the range for stable critical points P4 and P5. Specifically, we selected the same initial conditions and $\delta_1$ for all of them and different $\delta_2$, to show that as far as the chosen values are within the range, they all begin from the same points and approaches the attractors P4 or P5. If we would select other values for stability parameters within the range of stable critical point, still expecting to get the same behavior for the statefinder and effective EoS parameter.

\section{Cosmological Test}
We now examine our model with the observational data using the following cosmological tests \cite{f1}--\cite{Liske}.

\subsection{CPL, CRD and our model}

Following~\cite{f8}, in Chevallier-Polarski-Linder (CPL) parametrization model one can use linearly approximated EoS parameter,
\begin{eqnarray}\label{hdot1}
\omega_{cpl}\approx \omega_0-\frac{d\omega_{cpl}}{da}(a-1)=\omega_0+\omega_1\frac{z}{1+z},
\end{eqnarray}
where $\omega_0$ is current value of the EoS and $\omega_1=-\frac{d\omega_{cpl}}{da}$ is its running factor. Using the above equation we can find
the following equation for Hubble parameter,
\begin{equation}
\frac{H(z)^{2}}{H^{2}_{0}} =\Omega_m(1+z)^3+(1-\Omega_m)(1+z)^{3(1+\omega_0+\omega_1)}\times  \exp{\left[-3\omega_1(\frac{z}{1+z})\right]}\cdot \label{Hr}
 \end{equation}
In CPL model the parametrization is fitted for different values of $\omega_0$, $\omega_1$ and $\Omega_m$.

On the other hand, the CRD can be extracted from
\begin{eqnarray}\label{dotz}
\dot{z}=(1+z)H_0-H(z),
\end{eqnarray}
which is known as Mc Vittie equation. This equation immediately leads to velocity drift
\begin{eqnarray}\label{vdrift}
\dot{v}=cH_0-\frac{cH(z)}{1+z}.
\end{eqnarray}
By using equation (\ref{Hr}) in CPL model, the velocity drift with respect to the redshift can be obtained against observational data. In our model, from numerical computation one can obtain $H(z)$ which can be used to evaluate $\dot{v}$. To best fit the model for the parameters $\delta_1$ and $\delta_2$ and the initial conditions $x(0)$, $z(0)$, $w(0)$, we use the simulated data points for redshift
drift experiment generated by Monte Carlo simulations \cite{Liske} by employing the $\chi^2$ method. We constrain the parameters including the initial conditions by minimizing the $\chi^2$ function given as
\begin{equation}\label{chi20}
    \chi^2(\delta_1, \delta_2, x(0), z(0), w(0))=\sum_{i=1}^{8}\frac{[\dot{v}_i^{the}(z_i|\delta_1, \delta_2, x(0), z(0), w(0)) - \dot{v}_i^{obs}]^2}{\sigma_i^2},
\end{equation}
where the sum is over the three sets of
data (8 points) for redshift drift experiments \cite{Liske}. In relation (\ref{chi20}), $\dot{v}_i^{the}$ and $\dot{v}_i^{obs}$ are the velocity drift parameters obtained from our model and from observation, respectively, and $\sigma$ is the estimated error of the $\dot{v}_i^{obs}$. In our model the best fit values occur at $\delta_1=-10.5$, $\delta_2=3.5$, $x(0)=-0.6$, $z(0)=-0.2$, $w(0)=1.09$ with $\chi^2_{min}=1.27718546$. In Fig. 6, the velocity drift, $\dot{v}$, in our model is compared with the observational data for the obtained parameters and initial conditions using $\chi^2$ method in comparison with the CPL model and also $\Lambda$CDM model. As can be seen in our model for the best fitted stability parameters
 the model is in better agreement with the observational data in comparison with the CPL and $\Lambda$CDM models.\\

\begin{tabular*}{2.5 cm}{cc}
\includegraphics[scale=.4]{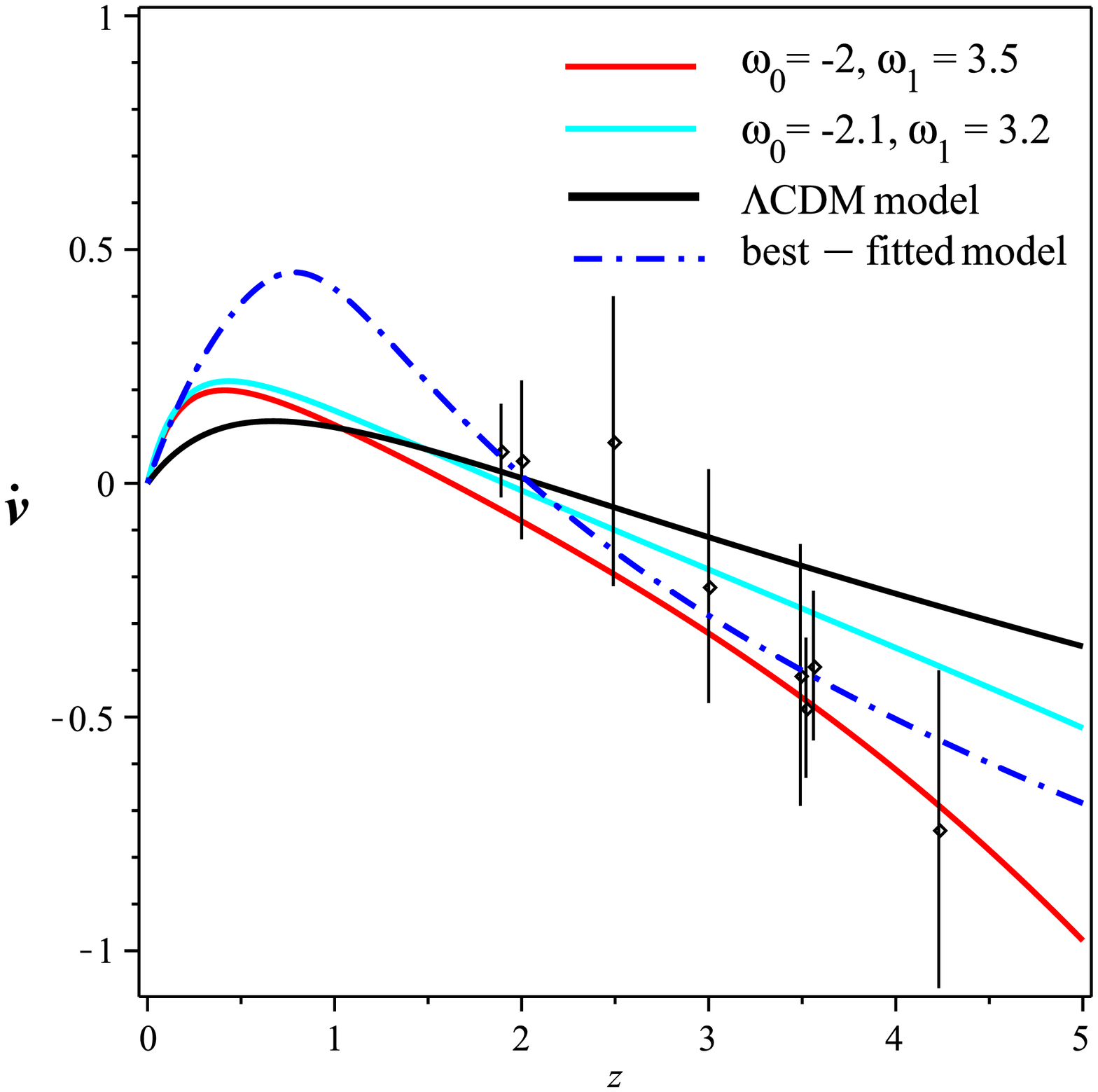}\hspace{0.1 cm}\\
Fig. 6: The graph of velocity drift plotted as function of redshift for CPL model, \\ our best-fitted model and $\Lambda$CDM . The stability parameter in our model: $\delta_1=-10.5$,\\
$\delta_2=3.5$. ICs. $x(0)=-0.6$, $z(0)=-0.2$, $w(0)=1.09$ and $\epsilon=0.7$.
\end{tabular*}\\

\subsection{The difference in the distance modulus, $\mu(z)$}

The difference between the absolute and
apparent luminosity of a distance object is given by, $\mu(z) = 25 + 5\log_{10}d_L(z)$ where the Luminosity distance quantity, $d_L(z)$ is given by
\begin{equation}\label{dl}
d_{L}(z)=(1+z)\int_0^z{\frac{dz'}{H(z')}}.
 \end{equation}
 In our model, from numerical computation one can obtain $H(z)$ which can be used to evaluate $\mu(z)$. To best fit the model for the parameters $\delta_1$ and $\delta_2$ and the initial conditions $x(0)$, $z(0)$, $w(0)$ with the most recent observational data, the Type Ia supernovea (SNe Ia), we employe the $\chi^2$ method. We constrain the model parameters including the initial conditions by minimizing the $\chi^2$ function given as
\begin{equation}\label{chi2}
    \chi^2_{SNe}(\delta_1, \delta_2, x(0), z(0), w(0))=\sum_{i=1}^{557}\frac{[\mu_i^{the}(z_i|\delta_1, \delta_2, x(0), z(0), w(0)) - \mu_i^{obs}]^2}{\sigma_i^2},
\end{equation}
where the sum is over the SNe Ia sample. In relation (\ref{chi2}), $\mu_i^{the}$ and $\mu_i^{obs}$ are the distance modulus parameters obtained from our model and from observation, respectively, and $\sigma$ is the estimated error of the $\mu_i^{obs}$.

For Gaussian distributed measurements, the likelihood
function is proportional to $e^{-\chi^2/2}$. In Fig. 7 we show the 2-dim likelihood distributions
for the parameters $\delta_1$ and $\delta_2$ as obtained from SNe Ia data. The parameters varies as  $1.1<\delta_1<3.1$ and  $-4<\delta_2<-5.6$ with the maximum likelihood at $\delta_1=2.1$, $\delta_2=-4.8$.

\begin{tabular*}{2.5 cm}{cc}
\includegraphics[scale=.4]{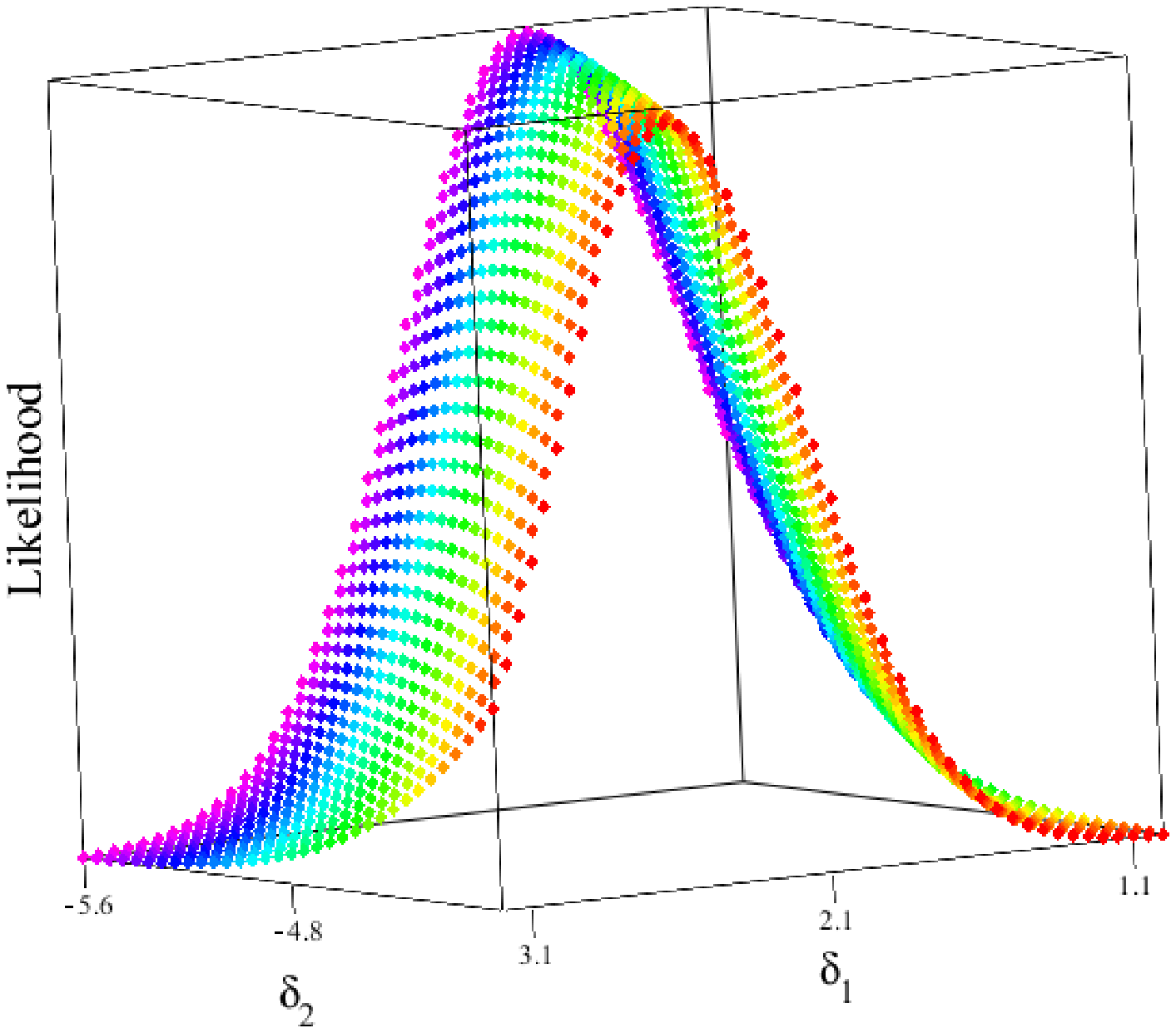}\hspace{0.1 cm}\\
Fig. 7: The graph of the 2-dim likelihood distribution for parameters $\delta_1$ and  $\delta_2$.\\
 ICs. $x(0)=0.6$,$z(0)=0.8$,$w(0)=1.1$,$d_L(0)=0$.\\
\end{tabular*}\\

In Fig. 8, the distance modulus, $\mu(z)$, in our model is compared with the observational data for the obtained parameters and initial conditions using $\chi^2$ method.\\

\begin{tabular*}{2.5 cm}{cc}
\includegraphics[scale=.4]{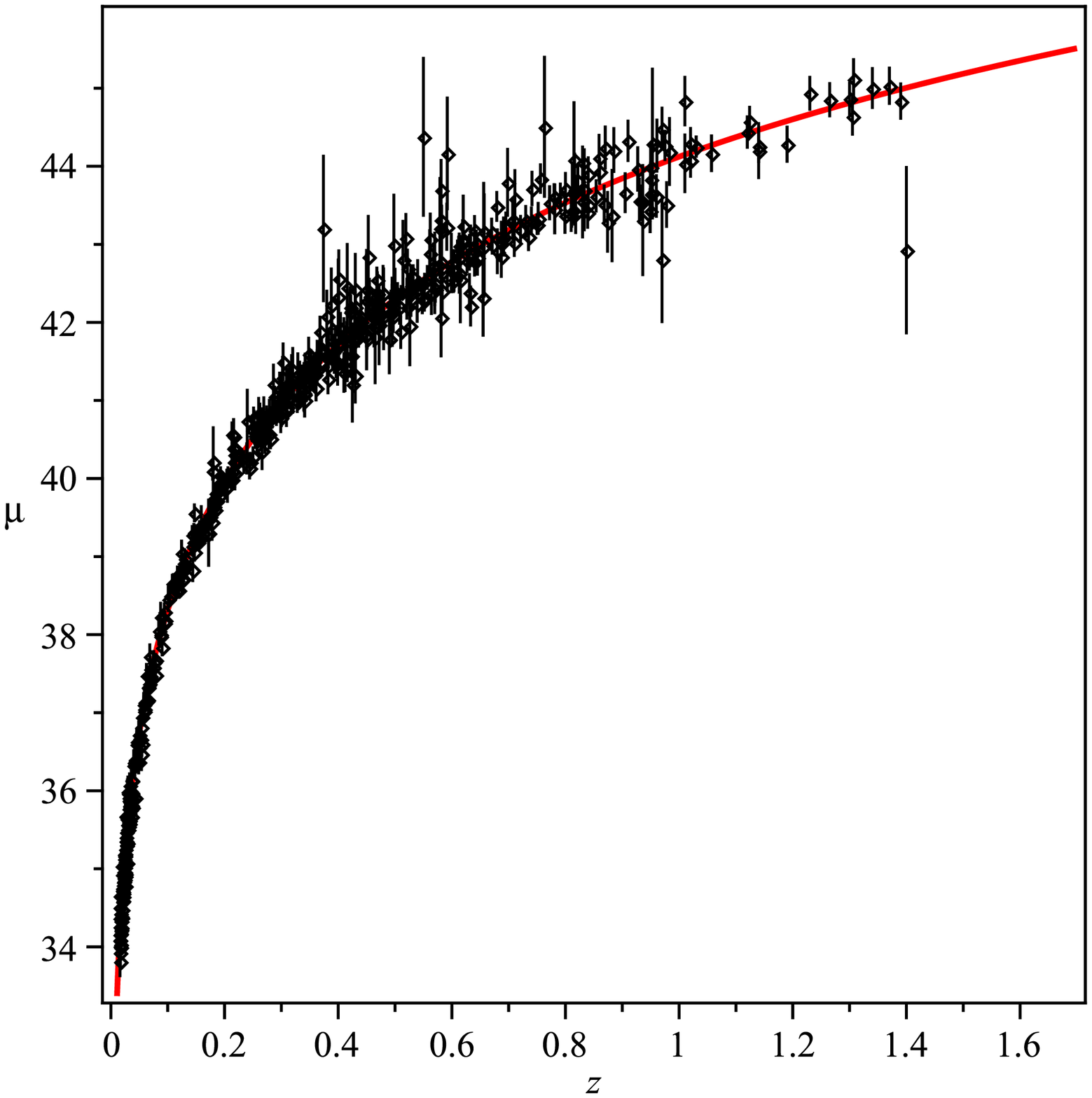}\hspace{0.1 cm}\\
Fig. 8: The graph of distance modulus $\mu(z)$ plotted as function of redshift, \\
for $\delta_1=2.1$, $\delta_2=-4.8$, $\epsilon=0.7$.
 ICs. $x(0)=0.6$,$z(0)=0.8$,$w(0)=1.1$,$d_L(0)=0$.\\
\end{tabular*}\\
In our model the best-fitted model parameters occur at $\delta_1=2.1$, $\delta_2=-4.8$, $x(0)=0.6$, $z(0)=0.8$, $w(0)=1.1$ with $\chi^2_{min}=541.6569729$.

\section{Summary and Conclusion}

This paper is designed to study the attractor solutions of tachyonic potential chameleon cosmology by stability analysis and making use of
 the 3-dim phase space of the theory. The model
 characterized by the scalar field $\phi$
 , the scalar potential $V(\phi)$, and the scalar function $f(\phi)$ nonminimally coupled to the matter lagrangian in the model. The phase
 space of the model is investigated using the scalar function $f(\phi)$ and potential $V(\phi)$ in exponential form. The stability analysis gives the corresponding conditions
for tracking attractor and determines the type
of the universe behavior in the past and future. It has been shown that there are two stable critical points in the 3-dim phase space
for the stability parameter $\delta_1$ and $\delta_2$. Together with the projection of the 3-dim phase space into 2-dim plane spaces, the
trajectories for different stability parameters and I.Cs. are shown in Figs. 1 and 2.

We then study the cosmological parameters such as effective EoS parameter, $\omega_{eff}$ and statefinder parameters
for the model in terms of the dynamical variables introduced in the stability section. It shows that the EoS parameter, $\omega_{eff}$, crosses the cosmological divide line in near future and become tangent to it in a stable state in far future. From Fig. 3, the universe starts from unstable state in the past with $\omega_{eff}=0$ and finally tends to the stable state in far future with $\omega_{eff}=-1$. It is notable that for the stable critical points P4 and P5, the squared sound speed, $c_s^2$ becomes minus infinity as $\omega_{eff}$ tends to -1 in these stable states. The violation of the sound speed requirement to be positive is due to the property of the phantom states. From the statefinder graphs we see that in all scenarios with different stability parameters and initial conditions the universe begins from the standard cold dark energy (SCDM) unstable state and eventually tends to the stable state near SS state. It also shows that the universe with negative deceleration parameter is currently accelerating. To test our models, two cosmological tests, CRD, and the distance modulus $\mu(z)$ are performed to compare our model with the CPL parametrization model, standard $\Lambda$CDM model and also with the observational data. In both tests, by employing the $\chi^2$ method we obtain a satisfactory fit to the observational data. Comparing the best-fitted velocity drift in our model with CPL and $\Lambda$CDM models shows that our model for the redshift around 2 and 3.5 better matches the observational data. The best-fitted distance modulus $\mu(z)$ with the observational data, is also shown in Fig.8.

\end{document}